\documentclass[pre,aps,twocolumn,showpacs]{revtex4}
\usepackage{graphicx}
\begin{document}
\title{Unusual features of coarsening with detachment rates decreasing with
cluster mass}
\author{F. D. A. Aar\~ao Reis${}^{1,}$\footnote{Email address: reis@if.uff.br}
and R. B. Stinchcombe${}^{2,}$\footnote{E-mail address:
r.stinchcombe1@physics.ox.ac.uk}}
\affiliation{
${}^{1}$ Instituto de F\'\i sica, Universidade Federal Fluminense, Avenida
Litor\^anea s/n, 24210-340 Niter\'oi RJ, Brazil
${}^{2}$ Rudolf Peierls Centre for Theoretical Physics, Oxford University, 1
Keble Road, Oxford OX1 3NP\\
}
\date{\today}
\begin{abstract}
We study conserved one-dimensional models of particle diffusion, attachment
and detachment from clusters, where the detachment rates decrease with
increasing cluster size as $\gamma (m)\sim m^{-k}$, $k>0$. Heuristic scaling
arguments based on random walk properties show that the typical cluster size
scales as ${\left( t/\ln{t}\right)}^z$, with $z=1/\left( k+2\right)$. The
coarsening of neighboring clusters is characterized by initial symmetric flux
of particles between them followed by an effectively assymmetric flux due to
the unbalanced detachement rates, which leads to the above logarithmic
corrections. Small clusters have densities of order $t^{-mz\left( 1\right)}$,
with $z\left( 1\right) = k/\left( k+2\right)$. Thus, for $k<1$, the small
clusters (mass of order unity) are statistically dominant and the average
cluster size does not scale as the size of typically large clusters does. We
also solve the Master equation of the model under an independent interval
approximation, which yields cluster distributions and
exponent relations and gives the correct dominant
coarsening exponent after suitable changes to incorporate effects of
correlations. The coarsening of typical large clusters is
described by the distribution 
$P_t\left( m\right) \sim 1/t^y f\left( m/t^z\right)$, with $y=2z$.
All results are confirmed by simulation, which also illustrates the unusual
features of cluster size distributions, with a power law decay for small masses
and a negatively skewed peak in the scaling region. The detachment rates
considered here can apply in the presence of strong attractive interactions, and
recent applications suggest that even more  rapid rate decays are also
physically realistic.
\end{abstract}

\maketitle

\section{Introduction}

Domain growth in far from equilibrium conditions is observed in phase separation
of mixtures, dynamics of glasses and island coarsening during or after
deposition of a thin film, among other systems
\cite{bray94,ritort,mevansreview,robin}. This motivated the proposal of many
statistical models which exhibit growth laws for the typical domain size in
the form $l\sim t^z$, where $z$ is a coarsening exponent \cite{mevansreview}. 
For instance, when a system is quenched from an homogeneous phase into a
broken-symmetry phase, two universality classes are frequently found, one of
them of curvature driven (or diffusive) growth \cite{allen,ohta}, with
$z=1/2$, and the other of conserved scalar order parameter \cite{slyozov,huse},
with $z=1/3$. However, many model dynamics do not obey detailed balance and may
lead to domain growth with other power law forms or with anomalous coarsening,
in which with $l$ grows slower
than any power of time. A continuous range of coarsening exponents may
be obtained by tuning a single parameter in models with relatively simple
physical mechanisms, e. g. single particle exchange between clusters
\cite{bennaim,igloi}. On the other hand, anomalous coarsening is found in
certain models that mimic glassy behavior or phase separation
\cite{east,kafri} 
(such behaviour is also present in models with detailed
balance under certain conditions \cite{shore}). A range of coarsening behaviors
is also obtained experimentally, e. g. in recent works on shaken granular
systems (${\left( \log{t}\right)}^{1/2}$) \cite{granular}, separation of
mixtures of milk protein and amylopectin ($0.04\leq z\leq 0.2$) \cite{bont} and
air bubbles in foams ($0.2\leq z\leq 0.5$) \cite{schmitt}.
Despite the variety of possible scenarios which were already shown in the
literature, the study of simple models with normal or anomalous coarsening  is
still important because it may reveal the basic microscopic mechanisms that lead
to certain macroscopic behavior. Such basic studies may also help the 
development of more realistic models for a wide range of processes, such as
those in Ref. \protect\cite{mccoy}.

A class of models in which islands grow via particle diffusion, attachment and
detachment (Ostwald ripening) is very important in surface science because they
can explain many features of submonolayer or multilayer growth
\cite{cv,biehl,etb}. Even the one-dimensional models are important in this
field, both as a first step to understand realistic two-dimensional systems and
as models for growth of elongated islands \cite{gambardella,tokar,coarsen1}.
These one-dimensional
models may usually be mapped onto zero-range processes (ZRP), whose universal
and non-univeral properties were intensively studied in the last years
\cite{evanshanney,spitzer}. Here, we will analyze the coarsening process in a
class of conserved one-dimensional models with those mechanisms, as illustrated
in Fig. 1a. The mapping to a column problem, which is a ZRP, is shown in Fig.
1b. Isolated adatoms diffuse with unit 
rate and attachment occurs immediately
after a particle reaches the border of a cluster. 
We study here the case in which  the rate of
detachment from a cluster decreases 
with increasing
cluster size as an inverse power law of the form 
\begin{equation}
\gamma\left( m\right) = \gamma_0 / m^k ,
\label{defgamma}
\end{equation}
with $k>0$. We consider a very large lattice (infinite for practical purposes),
where a non-trivial, continuous coarsening process is observed if the system
begins in a completely random configuration.

This form of detachment rate could apply with some type of long-range attraction
between
the particles in a cluster \cite{ammi}. This mechanism may not be generic for
usual surface science
applications, but the form may nevertheless be a 
reasonable approximation for a range of cluster
sizes. Moreover, it may find applications in other fields, such as granular
systems, where rates with much faster decay ($\gamma\sim \exp{\left(
-m^2\right)}$) were already used to model real systems \cite{granular}. This is
an important motivation for this study, and additional support to this claim is
provided by some of its unusual features. First,
cluster growth shows features that resemble other ZRP with biased diffusion
\cite{godreche,grobkinsky} because there is a preferential flux from the small
to the large clusters, despite the model rules being completely symmetric. The
coarsening exponent is $z=1/(k+2)$, but there is a logarithmic correction to
the dominant power-law coarsening. Thus, as $k\to 0$, we obtain $z\to 1/2$,
instead of the value $z=1/3$ obtained with symmetric rules in Ref.
\protect\cite{coarsen1} (constant $\gamma$) and Refs.
\protect\cite{godreche,grobkinsky} (decreasing $\gamma$, but $\gamma(m)\to 1$
as $m\to\infty$). On the other hand, the logarithmic correction represents the
crossover from symmetric to effectively assymmetric particle flux which occurs
during the exchange of particles between neighboring clusters. Another
interesting feature is the difference between the scaling of the average
cluster size (all clusters) and the scaling of the typical size of large
clusters for $k<1$, due to the presence of high densities of small clusters
dominating that average. This contrasts to related models,
including those with deposition and/or fragmentation, whose relevant cluster
sizes are described by a single scaling relation. These features are accompanied 
by cluster size distribution with non-usual features, including a
high negative skewness near the typical growing size.

\begin{figure}[!h]
\includegraphics[clip,width=0.4\textwidth,angle=0]{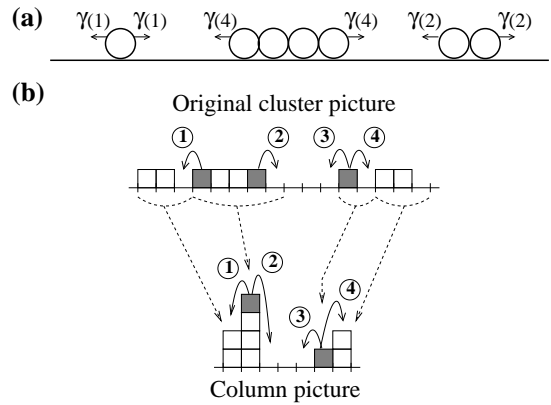}
\caption{(a) Illustration of the diffusion ($m=1$) and detachment
($m>1$) processes of the model, with the associated rates $\gamma(m)$. (b)
Examples of detachment processes (1,2) and diffusion processes (3,4) 
of shaded particles, 
in the original cluster picture and in the corresponding column
picture. Dashed lines show the correspondence between cluster+vacancy and a
column in the two pictures.}
\label{fig1}
\end{figure}

At this point, it is also important to recall the differences from previously
studied models with similar mechanisms. The case of constant detachment rate
(more precisely, $k=0$ and $\gamma_0\ll 1$) was considered in Refs.
\protect\cite{coarsen1,chamereis}, and shows a coarsening with exponent $z=1/3$
up to a characteristic time of order ${\gamma_0}^{-5/2}$. Models with
$\gamma(m)$ increasing with $m$ were also analyzed in previous work \cite{gama}
and have prospective application 
to island formation in heteroepitaxy,
particularly due to the possibility of changing the shape of the island size
distributions (from
monotonic to peaked ones) by tuning temperature or coverage. In those cases,
steady states could be attained in infinitely large lattices, but the present
model (decreasing $\gamma (m)$) shows a steady state only in a finite lattice.
The properties of this steady state can be exactly predicted from a
mapping onto a ZRP: for any rate of the form in Eq.(\ref{defgamma}), there is
condensation into a single cluster whose density tends
to $1$ as the lattice size increases \cite{evanshanney}.

Our results for the average cluster sizes, including the logarithmic corrections
to the dominant behavior, will be derived from a scaling theory presented in
Sec. \ref{scalingtheory} and will be confirmed by simulation data. In Sec.
\ref{iia}, we will write the Master equation of the process in an independent
interval approximation (IIA), and obtain some 
exponent relations.
However, because of its neglect of important correlations, 
some results of this IIA do not agree with the scaling ones. But, 
after some adjustment it is able to
predict the correct dominant coarsening exponent. The simulation results for
cluster size distributions are shown in Sec. \ref{distr}, which qualitatively
confirm the assymmetry predicted by the IIA and the proposed scaling relations
for small and for typically large clusters. Finally, in Sec. \ref{conclusion},
we present our conclusions.

\section{Scaling theory}
\label{scalingtheory}

\subsection{Basic definitions and coarsening with constant detachment rates}
\label{diffonly}

Here we review the heuristic scaling approach based on random  walk properties
used to predict the time evolution of the typical cluster size.
We consider the model with small mass-independent detachment rates, i. e.
$\gamma (m) =\gamma_0 \ll 1$ for $m\geq 2$, while $\gamma (1) =1$ (free particle
diffusion). These
arguments were formerly presented in Ref. \protect\cite{coarsen1} and follow
similar lines of those applied to other ZRP in Refs.
\protect\cite{grobkinsky,evanshanney}.
We denote the typical cluster size as $M$, which must be understood as an
average over the largest (time-increasing) sizes which are statistically
relevant. This average excludes, for instance, clusters with size of order
$1$, even if their statistical weights are large.

\begin{figure}[!h]
\includegraphics[clip,width=0.4\textwidth,angle=0]{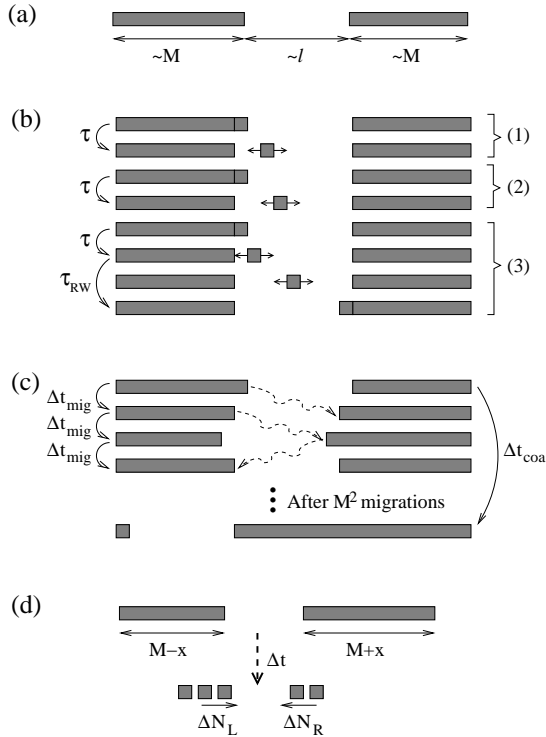}
\caption{(a) Configuration of two neighboring large clusters with typical size
$M$, separated by an empty region of size $l$. (b) Scheme with a sequence of
configurations during the migration of a particle from the left to the right
cluster. (c) Scheme with successive migrations of particles from one cluster to
the other, until the right cluster doubles its mass at the expense of the left
one. (d) Scheme with the number of particles ${\Delta N}_L$ and ${\Delta N}_R$
detached from neighboring clusters during a time interval $\Delta t$. From the
model rules, the smaller cluster (left) loses more particles than the larger
one (right).}
\label{fig2}
\end{figure}

Fig. 2a shows two neighboring clusters of size $M$ separated by a gap of size
$l=rM$, where
$r$ is related to the particle density (coverage) $\theta$ by
\begin{equation}
r\equiv \frac{\theta}{1-\theta} .
\label{defr}
\end{equation}
A characteristic time ${\Delta t}_{coa}$ is that in which such 
clusters exchange so many particles that one
of them approximately doubles its mass at the expense of the other. This 
time is estimated below.

The  time for 
detachment of a single particle from the edge of a cluster is of order
$\tau \sim 1/\gamma_0$. However, after detachment it is
much more probable for this particle to reattach to that cluster than to
diffuse to the other cluster. The probability of traveling a distance $l$
before going back to the original cluster is $1/l$, as
determined by the solution of "the gambler's ruin problem" \cite{feller} - see
also Refs. \protect\cite{coarsen1,grobkinsky}. This means that the particle
will detach and reattach to the original cluster a number of times of order
$l$ before migrating to the neighboring cluster. This is illustrated in Fig.
2b, where for simplicity only two unsucessful detachments (i. e.
detachment-reattachment), labeled (1) and (2),
were shown. Consequently, successful migration of a single particle from one
cluster to the other takes place after a time ${\Delta t}_{mig}$ given by
\begin{equation}
{\Delta t}_{mig} \sim \tau l \sim rM/\gamma_0 .
\label{tmig0}
\end{equation}
The additional time for random walk of the free particle, $\tau_{RW}$, is
negligible during coarsening.

The above reasoning implies that single particle exchange does not depend on the
current size of each cluster, but only on their separation $l$, which is kept
fixed during the process. This symmetric
random exchange is illustrated in Fig. 2c. After the migration time
$\tau_{mig}$, the size of each cluster increases or decreases by one unit
with equal probability. Thus, in order 
for the size of one of the clusters to increase from $M$ to $2M$ (and the size
of the other cluster
to decrease from $M$ to zero), the exchange of nearly $M^2$ particles is
necessary. Thus, the coarsening time is
\begin{equation}
{\Delta t}_{coa} \sim M^2 {\Delta t}_{mig} \sim rM^3/\gamma_0 .
\label{tcoa0}
\end{equation}
This gives a scaling equation
\begin{equation}
\frac{dM}{dt} \sim \frac{M}{{\Delta t}_{coa}} ,
\label{scalingeq0}
\end{equation}
from which we obtain
\begin{equation}
M \sim {\left( \frac{\gamma_0}{r} t\right) }^{1/3} .
\label{scalingdiffonly}
\end{equation}

Notice that the random walk of the free particle between the neighboring
clusters takes a time of order
\begin{equation}
\tau_{RW} \sim l^2 \sim {\left( rM\right) }^2 .
\label{trw}
\end{equation}
If $\tau_{RW} \ll \tau$, then during the successful migration time there will be
only one free particle between the clusters, as assumed above. Otherwise, if
$\tau_{RW} \sim \tau$, it is probable that two free particles meet, which leads
to the formation of an intermediate
cluster with those particles. Since the time necessary for the small
intermediate cluster to break is of the same order 
as the detachment rates from
the big clusters ($\tau\sim 1/\gamma_0$), the coarsening process ends. In this
situation, we have
$M\sim \frac{1}{r{\gamma_0}^{1/2}}$ for the average cluster size
\cite{coarsen1}.

\subsection{Coarsening with decreasing detachment rates}
\label{scalingcoars}

Here we extend the previous approach to the case of decreasing detachment rates
(Eq. \ref{defgamma}). In this case, the characteristic time for single particle
detachment from a typical cluster of size $M$ is
\begin{equation}
\tau (M) \sim \frac{1}{\gamma\left( M\right)} \sim \frac{M^k}{\gamma_0} ,
\label{deftau}
\end{equation}
where we used $\gamma_0 \sim 1$.

In contrast to the model with constant detachment rates (Sec. \ref{diffonly}),
here we observe that coarsening will not end in an
infinitely large lattice because, during the exchange of particles between
neighboring clusters, the time necessary to break the intermediate cluster is
of order $1$, which is much smaller than the detachment time. In a finite
lattice, this leads to condensation of a finite fraction of the particles 
into a single cluster (with the present rates, this fraction tends to $1$ as the
size increases) \cite{evanshanney}.

The time for successful migration from one cluster to the neighboring one is 
\begin{equation}
{\Delta t}_{mig} \sim \tau l \sim rM^{k+1}/\gamma_0 ,
\label{tmig}
\end{equation}
which now depends explicitly on the mass of the cluster from which it detached.
Detachment from large clusters is slower, thus there is a preferential flux of
particles from small to large neighboring clusters. Eq. (\ref{tcoa0}) is no
longer valid because the number of single particle exchanges
necessary for two clusters to coarsen is much smaller than $M^2$.
When the neighboring clusters
have nearly the same size, random exchange of particles takes place, but as
soon as the sizes are unbalanced the net flux becomes asymmetric.

The next step is to calculate the number of exchanged particles within a time
interval $\Delta t$ if the mass is unbalanced by an amount $x$, as shown in
Fig. 2d. The numbers of detached particles from the left and the right clusters
during that time are, respectively,
\begin{eqnarray}
{\Delta N}_L &\sim& \frac{\Delta t}{{\Delta t}_{mig}^{(LEFT)}} \sim
\gamma\left( M-x\right) \Delta t / \left( rM \right) , \nonumber \\
{\Delta N}_R &\sim& \frac{\Delta t}{{\Delta t}_{mig}^{(RIGHT)}} \sim
\gamma\left( M+x\right) \Delta t / \left( rM \right) .
\label{deltaN}
\end{eqnarray}
Consequently, the mass difference $x$ increases by
\begin{equation}
\Delta x = {\Delta N}_L - {\Delta N}_R \sim \frac{\gamma_0}{rM^{k+1}}
\left[ {\left( 1-\frac{x}{M} \right)}^{-k} - {\left( 1+\frac{x}{M}
\right)}^{-k} \right] {\Delta t} 
\label{deltax}
\end{equation}
within time $\Delta t$.

The time for a net flux of a fixed mass $\Delta x$ decreases as $x$
increases, which means slow coarsening for clusters of nearly the same size and
rapid coarsening with one big and one small cluster. Transfer of unit mass
($\Delta x=1$) takes place in a time of order
\begin{equation}
{\Delta t}_1 \sim \frac{rM^{k+1}}{\gamma_0} 
{\left[ {\left( 1-\frac{x}{M} \right)}^{-k} - {\left( 1+\frac{x}{M}
\right)}^{-k} \right]}^{-1} 
\label{deltat1}
\end{equation}
and the coarsening time is
\begin{equation}
{\Delta t}_{coa} = \sum_{x=1}^{x=M}{{\Delta t}_1} \sim \frac{rM^{k+2}}{\gamma_0}
\int_{1/M}^{1}{\frac{du}{{\left( 1-u\right)}^{-k} - {\left( 1+u\right)}^{-k}}}
\label{tcoa1}
\end{equation}

For typical masses $M\gg 1$, the integral in Eq. (\ref{tcoa1}) is dominated by
$u\ll 1$, where ${\left( 1-u\right)}^{-k} - {\left( 1+u\right)}^{-k} \approx 2ku
+ {\cal O}\left( u^3\right)$. Since we consider $k\sim 1$, we obtain
\begin{equation}
{\Delta t}_{coa} \sim \frac{r}{\gamma_0} M^{k+2} \ln{M} .
\label{tcoa3}
\end{equation}
Notice that $u\ll 1$ in Eq. (\ref{tcoa1}), which leads to the logarithmic
correction in Eq. (\ref{tcoa3}), physically corresponds to the regime of
symmetric particle exchange, i. e. neighboring clusters with approximately the
same size. Similar arguments were used to calculate coarsening times in Ref.
\protect\cite{torok}.
Since $k>0$, we observe that ${\Delta t}_{coa}$ is always larger than the
time for random walk between the clusters, given by Eq. (\ref{trw}), thus
particle detachment is always the leading contribution to the coarsening time
of large clusters.

Substituting Eq. (\ref{tcoa3}) in the scaling equation (\ref{scalingeq0}), we
obtain
\begin{equation}
M\sim {\left[ \frac{\gamma_0}{r} \frac{t}{\ln{t}}\right]}^z ,\qquad z =
\frac{1}{k+2} .
\label{Mt}
\end{equation}

In order to test these predictions, we performed numerical simulations of the
model for several values of
$k$ in the range $[0.25,3]$, with coverages $\theta =0.8$, in lattices of sizes
from $L=8192$ to $L=32768$, so that finite-size effects are negligible.
Simulations for some
smaller coverages were also performed, but the coarsening process usually takes
place at much longer times. The average cluster size
${\langle m\rangle}$ was obtained from at least 100 configurations for each $K$,
up to times of order $t={10}^6$.

Estimates of the exponent $z$ are usually obtained from extrapolation of
effective exponents calculated from ${\langle m\rangle}\left( t\right)$.
Without accounting for logarithmic corrections in Eq. (\ref{Mt}), we define the
effective exponents as
\begin{equation}
z_{eff,1} = \frac{ \ln{\left[ \langle m\rangle\left( t\right) / \langle m\rangle
\left( t-\delta t\right) \right] }}{ \ln{\left[ t/\left( t-\delta
t\right)\right]} } ,
\label{zeff1}
\end{equation}
with fixed $\delta t$.
On the other hand, in order to account for the logarithmic corrections in Eq.
(\ref{Mt}), the effective exponents must be defined as
\begin{equation}
z_{eff,2} = \frac{ \ln{ \left[ \langle m\rangle\left( t\right) /
\langle m\rangle \left( t-\delta t\right) \right] } }{
\ln{ \left[ \left( t/\ln{t}\right) / \left( \left( t-\delta t\right) /
\ln{\left( t-\delta t\right)} \right) \right] } } .
\label{zeff2}
\end{equation}

$z_{eff,1}$ is plotted in Fig. 3a as a function of $1/t$ for $k=3$, $k=2$, $k=1$
and $k=0.25$, and $z_{eff,2}$ is plotted in Fig. 3b for the same values of $k$.
Predicted asymptotic values $z=1/\left( k+2\right)$ (Eq. \ref{Mt}) are $0.2$,
$0.25$, $0.333$ and $0.444$, respectively.
For all $k\geq 1$, we observe that convergence to the asymptotic $z$ (as $1/t\to
0$) is faster with $z_{eff,2}$. This justifies the theoretically predicted
logarithmic corrections.

However, for $k=0.25$ we observe that both $z_{eff,1}$ and $z_{eff,2}$ converge
to $z\approx 0.12$, which is very far from the predicted value of Eq.
(\ref{Mt}). In Sec. \ref{roleisolated}, we will show that for
$k<1$ the coarsening exponent for ${\langle m\rangle}$ is actually different
from $z=1/(k+2)$ due to the large density of isolated particles. Thus 
${\langle m\rangle}$ is very different from $M$, which represents the
typical size of large, increasing clusters. However, we will show
that $M$ still coarsens with the exponent given by Eq. (\ref{Mt}).

\begin{figure}[!h]
\includegraphics[clip,width=0.4\textwidth,angle=0]{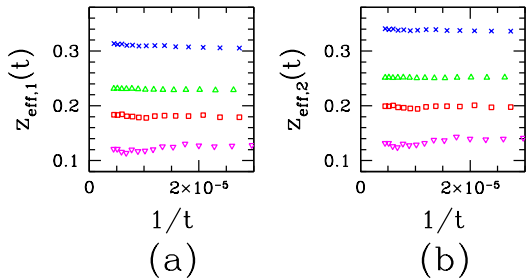}
\caption{(Color online) Effective exponents $z_{eff,1}$ (a) and $z_{eff,2}$ (b)
of the average cluster size (average over all clusters) 
as a function of inverse time, with coverage $\theta
=0.8$: $k=3$ (squares), $k=2$ (up triangles), $k=1$ (crosses), and $k=0.25$
(down triangles).}
\label{fig4}
\end{figure}

\subsection{The role of isolated particles}
\label{roleisolated}

The successful detachment of a particle from a cluster, which allows the
migration to the neighboring one, takes place after a time interval given by
Eq. (\ref{tmig}). However, this time measures the average residence time of the
particle attached to the original cluster. The total time of migration of a
single particle has to include the random walk time between the neighboring
clusters, which is given by Eq. (\ref{trw}).

If $t_{mig}>\tau_{RW}$, then the random walk is rapid, thus it is very rare to
observe a single free particle between
any pair of clusters and even rarer to observe two. This condition is satisfied
when $k>1$.
Fig. 4a shows some snapshots of the simulation for $k=2$, which confirm this
behavior. Thus, the large clusters with mass of order $M$ are statistically
dominant, i. e. $M$ actually represents the average cluster mass among all
clusters, which we denote by $\langle m\rangle$.

\begin{figure}[!h]
\includegraphics[clip,width=0.4\textwidth,angle=0]{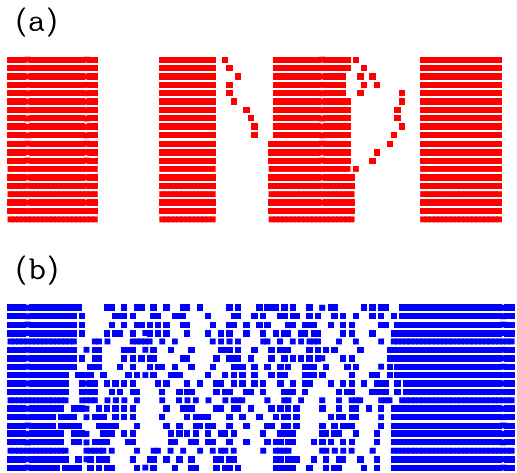}
\caption{(Color online) Sequences of configurations (from top to bottom) of a
certain region of the lattice for (a) $k=2$ and (b) $k=0.5$. In both cases, the
coverage is $\theta =0.6$ and snapshots are separated by a time interval $10$
(simulation times are of order ${10}^4$).}
\label{fig5}
\end{figure}

On the other hand, if $k<1$, a large time is spent in the random walk between
neighboring clusters. During this time, the successful detachment of other
particles is possible (we recall that intermediate small clusters rapidly break
for $\gamma_0 \sim 1$). This is illustrated for $k=1/2$ in the snapshots of
Fig. 4b. The number of free particles during $\tau_{RW}$ in the region between
two large clusters is of order
\begin{equation}
N_1 \sim \tau_{RW}/t_{mig} \sim M^{1-k}
\label {nfree}
\end{equation}
and the corresponding density of free particles is
\begin{equation}
\rho_1 \sim N_1 /M \sim M^{-k} .
\label {rho1M}
\end{equation}

However, the density of large clusters, whose typical mass is $M$, varies as
\begin{equation}
\rho_{large} \sim 1/M .
\label{rholarge}
\end{equation}
This means that the free particles (or small clusters formed by their
attachment) are statistically dominant for $k<1$. In this situation, $M$
represents the average size of large clusters, but not the average size among
all clusters, which is $\langle m\rangle$.

For $k>1$, Eq.
(\ref{rho1M}) is also valid as a density averaged in space and time (during
most of the time, there is no free particle between the neighboring clusters),
thus large clusters of size $M$ are statistically dominant and $\langle
m\rangle \approx M$.

These results do not invalidate the arguments of Sec. \ref{diffonly} for the
scaling of
$M$, which is still expected to follow Eq. (\ref{Mt}) for $k<1$. The
average cluster size calculated among all
clusters, including free particles, is obtained from an average in the region
between two large clusters:
\begin{equation}
\langle m\rangle \sim \frac{1\cdot N_1 + M\cdot 1}{N_1+1} \sim
\frac{M}{1+M^{1-k}} .
\label{mglobal}
\end{equation}
With $k<1$, this global average scales as
\begin{equation}
\langle m\rangle \sim M^k \sim {\left( \frac{t}{\ln{t}} \right)}^{z_G} ,
\qquad z_G = \frac{k}{k+2} \qquad (k<1).
\label{zG}
\end{equation}

This explains the discrepancies in the numerical estimates of coarsening
exponents for $k<1$ (Sec. \ref{scalingcoars}). For instance, for $k=0.25$, Eq.
(\ref{zG}) predicts $z_G=0.111$, which is consistent with the trend of the data
in Figs. 3a and 3b.

In order to test the predicted scaling of $M(t)$, we calculated numerically
average cluster sizes from contributions of large clusters only (masses $m>12$
for $k=0.5$, $m>25$ for $k=0.25$). Corresponding effective exponents are
defined as 
\begin{equation}
z_{eff,3} = \frac{ \ln{\left[ M\left( t\right) / M \left( t-\delta t\right)
\right] }}{ \ln{\left[ \left( t/\ln{t}\right) /\left( \left( t-\delta
t\right) / \ln{\left( t-\delta t\right)} \right) \right]} } .
\label{zeff3}
\end{equation}

\begin{figure}[!h]
\includegraphics[clip,width=0.4\textwidth,angle=0]{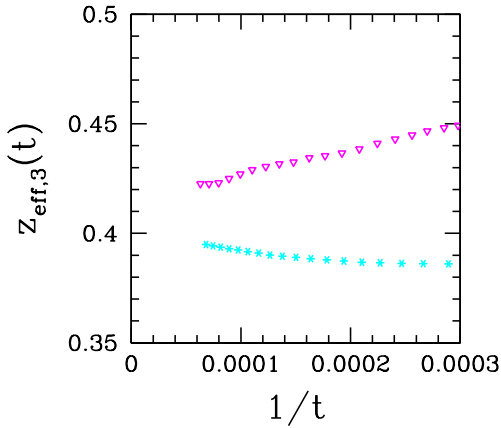}
\caption{(Color online) Effective exponents $z_{eff,3}$ for average cluster
sizes excluding small clusters, with coverage $\theta =0.8$: $k=0.5$
(asterisks) and $k=0.25$ (down triangles).}
\label{fig6}
\end{figure}

$z_{eff,3}$ is shown in Fig. 5 as a function of $1/t$ for $k=0.25$ and $k=0.5$.
Good agreement with the predicted asymptotic value $z=0.4$ for $k=0.5$ is
obtained. For $k=0.25$, the trend of $z_{eff,3}$ as $1/t\to 0$ is not
consistent with the predicted value $0.444$, which is probably due to
corrections to scaling. In both cases, effective exponents not accounting for
the logarithmic corrections (similarly to $z_{eff,1}$ - Eq. \ref{zeff1}) show
larger discrepancies from the theoretically predicted values of $z$.

Additional support to our theoretical predictions is provided by the numerical
study of the scaling of the density of free particles. From Eqs. (\ref{Mt}) and
(\ref{nfree}), we obtain
\begin{equation}
\rho_1 \sim {\left(\frac{t}{\ln{t}}\right)}^{-z\left( 1\right)} ,
\label {rhofree}
\end{equation}
with
\begin{equation}
z\left( 1\right) =  \frac{k}{k+2} .
\label{z1}
\end{equation}
(i. e. $z\left( 1\right) =z_G$ for $k<1$).
In Fig. 6 we show
${\left[ t/\ln{\left( t\right)}\right]}^{z\left( 1\right)}\rho_1$
versus $1/t$ for
$k=2$, $k=1$ and $k=1/2$, using the exponents $z(1)$ given by Eq.
(\ref{z1}). The convergence of that ratio to finite non-zero values as
$t\to\infty$ confirms the expected scaling.

\begin{figure}[!h]
\includegraphics[clip,width=0.4\textwidth,angle=0]{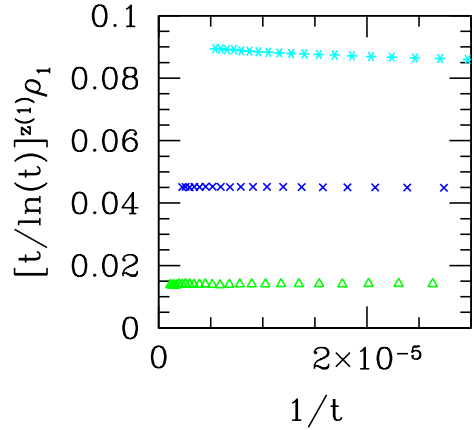}
\caption{(Color online) Simulation results for the scaled density of isolated
particles as a function of inverse time, with coverage $\theta
=0.8$: $k=2$ (triangles), $k=1$ (crosses) and $k=0.5$ (asterisks).}
\label{fig7}
\end{figure}

The densities of other small clusters can be obtained from $\rho_1$ by observing
that they have high detachement rates and, consequently, they may be viewed as
a set of nearly free particles at consecutive lattice sites. This reasoning
gives the density of clusters of size $m$, for $m\sim 1$ as
\begin{equation}
\rho_m \sim {\left( \rho_1\right)}^m \sim 
{\left(\frac{t}{\ln{t}}\right)}^{-z\left( m\right)} ,
\label{rhom}
\end{equation}
with
\begin{equation}
z\left( m\right) =  m z\left( 1\right) .
\label{zm}
\end{equation}
Simulations also confirm this result for  small clusters, such as $m=2$ and
$m=3$, for
several values of $k$.

\section{Relation to other models}

Our model may be mapped onto a column problem which clearly shows that it is a
ZRP. A cluster of length $m$ in the original problem and the vacant
site at its right side is represented by a column of mass $m$ in this new
picture. The mapping is illustrated in Fig. 1b. Sets of $n$ consecutive
vacancies in the original problem are represented by $n-1$ vacant columns in
the new picture. The detachment and diffusion processes correspond to hopping
of a particle from a column to the neighboring one. The mass-dependence of
detachment rates is translated into mass-dependent hopping rates
$\gamma(m)=2\epsilon(m)$
in order to account for the detachment in two edges of each cluster, each one
with rate $\epsilon(m)$.

In a finite lattice, condensation of a finite fraction of the mass in a single
cluster is expected for all densities if $\gamma(m)\to 0$ for $m\to\infty$.
Moreover, the density of particles out of the condensate decreases as
$L\to\infty$, as explained in Ref. \protect\cite{evanshanney}. This
is the case of our model, and our simulations in small lattices confirm those
steady state features.

However, while steady state properties of ZRP can be analytically calculated,
the coarsening process in infinitely large lattices is much more difficult to
predict. That is the reason why we use scaling approaches, simulation and
analytical tools based on suitable approximations (Sec. \ref{iia}) to study
coarsening of our model.

Comparison with related models is interesting at this point. Godr\'eche
\cite{godreche} and Gro$\beta$kinsky et al \cite{grobkinsky} analyzed the ZRP
with hopping rates $\gamma(m) = 1+b/m$ using heuristic arguments similar to
ours (see also review in Ref. \protect\cite{evanshanney}). They considered the
cases of symmetric and asymmetric hopping rates, which lead to average cluster
size scaling as $\langle m\rangle \sim t^{1/3}$ and $\langle m\rangle \sim
t^{1/2}$, respectively. The symmetric case is somehow equivalent to our model
with mass-independent detachment rates (Sec. \ref{diffonly}), since both have
constant and nonzero $\gamma (m)$ for $m\to\infty$ (very large clusters).

However, it is important to notice that our model with $k\to 0$, i. e. with very
weak mass-dependence of hopping rates, has $z\to 1/2$, in contrast to $z=1/3$
which characterizes constant detachment rates. Both models consider symmetric
hopping rates, but the asymmetric flux of mass between the neighboring clusters
in our model is always present and is responsible for the faster coarsening,
even if $k$ is very small. In other words, coarsening in the model with $k\to
0$ is very different from that with $k=0$.

On the other hand, we note that $z=1/3$ is obtained in our model for $k=1$. In
this case, the detachment rates decreasing with cluster size tend to make the
coarsening slower, and balances the effect of the asymmetric particle flux
between clusters, which favors faster coarsening. For $k>1$, mechanisms
favoring slow coarsening are stronger, thus $z<1/3$. For $k<1$, mechanisms
favoring fast coarsening are stronger, thus $z>1/3$. However, both mechanisms
are absent in the model with constant $\gamma$ and in the model of
Gro$\beta$kinsky et al \cite{grobkinsky}, both having $z=1/3$. 

The above discussion leads to the the conclusion that the same exponents may be 
obtained with different microscopic dynamics, while apparently similar dynamics
may lead to very different coarsening exponents.  It is important that such
features are considered if one aims to model real systems by ZRP or similar
models.

\section{Independent interval approximation}
\label{iia}

\subsection{General formulation}

The full analytic description of systems with stochastic processes such as those
of our model is provided by the Master equation, which is most easily written
in the column picture of Fig. 1b. Previously, this approach was used to study
the (exact) steady states of related models which correspond to ZRP
\cite{coarsen1,gama} and the coarsening in models with increasing number of
particles due to deposition processes \cite{coarsen2}.

The description of the present model is simplified by the fact that the process
conserves
the total particle numbers $N$. Thus, using periodic boundary conditions and a
total number of sites $L$ (lattice length in the original cluster picture), and
denoting
by $N_t(m)$ the total number of clusters of size $m$ ($\geq 1$) at time $t$, it
follows that (i) $N= \sum_{m=1}^{\infty}{mN_t(m)}$, (ii) the number of spacers
in the column picture
is $\sum_{m=1}^{\infty}{N_t(m)}$, and (iii) $N_t(m)$ equals the number of
columns of size $m$, for $m>0$. Hence, denoting by $N_t(0)$ the number of
columns of size
zero, we have $\sum_{m=0}^{\infty}{N_t(m)} = L-N\equiv L(1-\theta)$ (the last
step defining the coverage $\theta$ in the original picture). Thus the total
number of columns (including those of size zero) is a constant $L_c=L-N$. The
density in the column picture is $\rho_c\equiv N/L_c = r$ (Eq. \ref{defr}).

The system configuration can be specified by the ordered set of 
numbers of particles in each of the columns in succession: $\left( m_1,m_2
\dots m_{L_c}\right) = \{ m_i\}$.
The probability $P_t\{ m_i\}$ at time $t$ of the configuration
$\{ m_i\}$ changes by in and out processes. Collecting the effects of all such
processes in a time step $t\to t+1$ (see e. g. Refs.
\protect\cite{coarsen1,gama}) gives the full Master equation
\begin{eqnarray}
P_{t+1} \{ m_i\} - P_t \{ m_i\} = \nonumber\\
\sum_{l=1}^L
[ \gamma\left( m_{l-1}+1\right)
P_t \left( \dots m_{l-1}+1 , m_l-1 \dots\right) \nonumber\\
 + \gamma\left( m_{l+1}+1\right)
P_t \left( \dots m_{l}-1 , m_{l+1}+1 \dots\right) \nonumber\\
 - 2 \gamma\left( m_l\right)
P_t \{ m_i\} ]  \theta\left( m_l\right) .
\label{mastergeral}
\end{eqnarray}
The theta function above (zero for $m\leq 0$, otherwise unity) is actually
redundant as $P_t\left( \dots m-1 \dots\right)$ and $\gamma(m)$ vanish for
$m\leq 0$.

The Independent Interval Approximation (IIA) assumes 
that the configuration probability  $P_t\{ m_i\}$ can be factorised
as $\prod_{l=1}^{L-N}{P_{t,l}\left( m_l\right)}$.
That leads to a reduced form of
the Master equation in which cluster-cluster correlations are neglected:
\begin{equation}
P_{t+1,l}\left( m\right) - P_{t,l}\left( m\right) = {\cal
A}_t\left( m+1,l\right) - {\cal A}_t\left( m,l\right) ,
\label{master1}
\end{equation}
where
\begin{eqnarray}
{\cal A}_t \left( m,l\right) &\equiv& P_{t,l}\left( m\right)
\gamma\left( m\right)
\Theta\left( m-1\right) + \delta_{m,1} \gamma\left( 1\right)
P_{t,l}\left( 1\right) \nonumber \\
&& - \left[ P_{t,l}\left( m-1\right)
\Theta\left( m-1\right) + \delta_{m,1} P_{t,l}\left( 0\right) \right] {\cal
J}\left( l\right) \nonumber \\
&=& \Theta \left( m\right) A_t\left( m,l\right) ,
\label{defA}
\end{eqnarray}
with
\begin{equation}
A_t\left( m,l\right) = P_{t,l}\left( m\right) \gamma\left( m\right) -
P_{t,l}\left( m-1\right) {\cal J}\left( l\right) ,
\label{defA1}
\end{equation}
and
\begin{equation}
{\cal J}\left( l\right) \equiv \sum_{m=1}^{\infty}{\frac{1}{2}
\gamma\left( m\right) \left[ P_{t,l-1}\left( m\right) +
P_{t,l+1}\left( m\right) \right] } .
\label{defJ}
\end{equation}
In Eq. (\ref{defA}), the Theta function $\Theta (m)$ is zero for $m\leq 0$,
otherwise it is unity.
A further reduction results from neglecting dependences on the column label $l$,
so $P_{tl} \left( m\right)$ becomes $P_{t} \left( m\right)$. 
This form of IIA gives
\begin{equation}
P_{t+1}\left( m\right) - P_{t}\left( m\right) = A_t\left( m+1\right) - 
A_t\left( m\right) \Theta \left( m\right) ,
\label{masteriia}
\end{equation}
where
\begin{equation}
A_t\left( m\right) = P_t\left( m\right) \gamma\left( m\right) - \Gamma_t
P_{t}\left( m-1\right) , m\geq 1 ,
\label{Aiia}
\end{equation}
and
\begin{equation}
\Gamma_t = \sum_{m-1}^{\infty}{\gamma\left( m\right) P_t\left( m\right)} .
\label{Gamma}
\end{equation}

A useful result from the IIA equation (\ref{masteriia})
for large masses $m$ is
\begin{equation}
\sum_{m'=m}^{\infty}{P_t\left( m'\right) - P_{t+1}\left( m'\right) } = 
A_t\left( m\right) .
\label{difP}
\end{equation}

Hereafter we consider the mass-dependent rates in Eq. (\ref{defgamma}) for
$m\geq 1$. Unless otherwise stated, we will proceed with developments without
dependence on column label $l$, i. e. starting from Eqs. (\ref{masteriia}),
(\ref{Aiia}), (\ref{Gamma}) and (\ref{difP}), with $\sum_{m=0}^{\infty}
{P_t\left( m\right)} = 1$. Notice that $P_t\left( m\right)$ here differs from
the density $\rho_m$ in Sec. (\ref{scalingtheory}) by a constant factor
$1-\theta$ due to the different lattice lengths used to normalize probabilities
in different pictures.

\subsection{Scaling characteristics}
\label{scalingIIA}

The late time coarsening of large characteristic masses is expected to be
described by 
\begin{equation}
P_t\left( m\right) \sim \frac{1}{t^{y}} f\left( \frac{m}{t^z}\right) .
\label{scalingP}
\end{equation}
The exponents $y$ and $z$ depend on $k$, and $y$ need not equal $z$ because the
large masses need not dominate the normalisation sums, as shown in Sec.
\ref{roleisolated}. The region of the cluster size distribution where the
scaling equation (\ref{scalingP}) applies and masses are of order $t^z$ is
hereafter called region S. 

For small $m$, we have
\begin{equation}
P_t\left( m\right) \sim t^{-z\left( m\right)} , m\ll t^z ,
\label{Psmall}
\end{equation}
where $z(m)$ is defined consistently with Eq. (\ref{rhom}). This region is
hereafter denoted as A.

Finally, $P_t\left( 0\right)$ may strongly contribute to normalisation sums
because, as coarsening continues and $P(m)$ at small $m$ decreases, $P_t (0)$
will approach $1$. So, at late times,
\begin{equation}
1-P_t\left( 0\right) \sim t^{-z_A} ,
\label{P0}
\end{equation}
which defines $z_A$.

\subsection{Direct results for small clusters}
\label{smallclusteriia}

The IIA equations and the above definitions and properties directly lead to some
results for small $m$ and large times. This is a quasistatic situation in which
probabilities slowly vary in time, thus the left hand side (LHS) of Eq.
(\ref{masteriia}) is negligible. Since Eq. (\ref{masteriia}) is valid for all
$m\geq 0$ this leads to $A_t(m)\sim 0$, and Eq. (\ref{Aiia}) leads to
\begin{equation}
P_t\left( m\right) \sim P_t\left( 0\right) {\Gamma_t}^m {\left( m!\right)}^k .
\label{ptm}
\end{equation} 
Since $P_t\left( 0\right)\sim 1$, this yelds the form (\ref{Psmall}) and
confirms the relation (\ref{zm}) among the coarsening exponents of small $m$
given that
\begin{equation}
\Gamma_t \propto t^{-z\left( 1\right)} .
\label{Gammat}
\end{equation}

The sizes of the terms on the LHS and on the right hand side of Eq.
(\ref{masteriia}) are respectively, for a given $m$, of order
$(d/dt)\left[ P_t\left( m\right)\right] \sim t^{-1-z\left( m\right)} =
t^{-1-mz\left( 1\right)}$ and $A_t\left( m+1\right) \sim 
t^{-z\left( m+1\right)} = t^{-\left( m+1\right) z\left( 1\right)}$. The
quasistatic assumption means that the former is negligible compared to the
latter quantity at long times, thus
\begin{equation}
z\left( 1\right) < 1 .
\label{z1m1}
\end{equation}
This result is also consistent with the scaling picture of Sec.
\ref{scalingtheory} and simulation results.

The sum in Eq. (\ref{Gamma}) can then be separated into the contributions from
the two regions, A and S. Eqs. (\ref{Psmall}) 
and (\ref{Gammat}) [with (\ref{zm})] apply to A and 
Eq. (\ref{scalingP}) applies to S, thus
\begin{eqnarray}
\Gamma_t &\sim & \sum_{m=1}^{m_0\left( t\right)}{\left[ \dots
t^{-mz\left( 1\right)}\gamma\left( m\right) \right]} +
\int_{m_0\left( t\right)}^{\infty}{m^{-k} t^{-y} f\left(
\frac{m}{t^z}\right) dm} \nonumber \\
&\sim & \dots t^{-z\left( 1\right)} +\dots + t^{-\left[ y
+\left( k-1\right)z\right]} ,
\label{Gammaserie}
\end{eqnarray}
with $1\ll m_0\left( t\right) \ll t^z$. Eq. (\ref{Gammaserie}) is consistent
with (\ref{Gammat}) if
\begin{equation}
z\left( 1\right) \leq y + \left( k-1\right)z .
\label{desig}
\end{equation}

Simulations strongly support Eqs. (\ref{Gammat}) and (\ref{Gammaserie}), as
well as (\ref{desig}) as an inequality (which is also consistent with the
scaling theory, as discussed below). This implies that the sum in $\Gamma_t$ is
dominated by the small $m$ region (actually by just the $m=1$ term). The result
(\ref{ptm}), which implies
$P_t\left( m\right)/{\left[ P_t\left( 1\right)\right]}^m = {\left( m
\right)}^k$, is also confirmed by simulation.

\subsection{Results for large clusters and exponents relations}
\label{exponentsiia}

Here we denote by $\sum_A$ and $\sum_S$ the summations with respect to $m$ over
regions A and S, respectively.

Consider the sum giving the density in the column picture
\begin{eqnarray}
\rho_c &=& \sum_{m=1}^\infty{mP_t\left( m\right)} = \nonumber\\
&&\sum_A{mP_t\left( m\right)}
+ \sum_S{mP_t\left( m\right)} = \nonumber\\
&&\dots t^{-z\left( 1\right)} +\dots
t^{-\left( y-2z\right)}
\label{rhoc}
\end{eqnarray}
(the sums being carried out in same way as those giving Eq. 
\ref{Gammaserie}).
Since $\rho_c$ is constant in time, this is consistent with
\begin{equation}
y = 2z .
\label{yz}
\end{equation}

Similarly, the density of clusters in the scaling region S is 
\begin{equation}
\sum_S{P_t\left( m\right)} \propto t^{-\left( y-z\right)} \propto t^{-z} ,
\label{somaS}
\end{equation}
where we used Eq. (\ref{yz}).

A further exponent relation follows from Eq. (\ref{P0}) and
\begin{eqnarray}
1-P_t\left( 0\right) &=& \sum_A{P_t\left( m\right)} + \sum_S{P_t\left( m\right)} =
\nonumber\\
&&\dots t^{-z\left( 1\right)} +\dots t^{-z} = \dots t^{z_A} ,
\label{P0AS}
\end{eqnarray}
which gives
\begin{equation}
z_A = min\{ z\left( 1\right) , z\} .
\label{zA}
\end{equation}
It turns out that the minumum here is $z(1)$ for $k<1$ and $z$ for $k\geq 1$,
where small and large clusters are respectively dominant (this was shown in Sec.
\ref{scalingtheory} and will be confirmed in the context of the IIA below).

These considerations warns us that there are several average masses, including
\begin{eqnarray}
\sum_S{mP_t\left( m\right) }/\sum_S{P_t\left( m\right) } \propto  \nonumber\\
t^z
\sum_{m=1}^{\infty}{mP_t\left( m\right) }/\sum_{m=1}^{\infty}
{P_t\left( m\right) }\propto t^{z_A} ,
\label{somas}
\end{eqnarray}
with $z_A$ given by Eq. (\ref{zA}).

Now consider the IIA Master equation in the form Eq. (\ref{difP}),
and the ansatz
for the scaling regime, Eq. (\ref{scalingP}). Replacing the time difference by
a derivative and the sum over $m$ by an integral, we have (also using Eqs.
\ref{defgamma}, \ref{Aiia} and \ref{Gammat})
\begin{eqnarray}
\int_m^\infty{dm' \frac{\partial}{\partial t}\left[ t^{-y} f\left(
\frac{m'}{t^z}\right)\right] } =\nonumber\\
 \left[ \dots m^{-k} - \dots
t^{-z\left( 1\right)} \left( 1-\frac{\partial}{\partial m} \right) \right]
t^{-y} f\left(\frac{m}{t^z}\right) .
\label{iiaS}
\end{eqnarray}
The leading order terms on the RHS cannot cancel, since they have different
dependences on $m$, so we can ignore the subdominant $\frac{\partial}{\partial
m}$ (which came from the $m-1$ argument). With $x\equiv m/t^z$ and $u=m'/t^z$,
the result is
\begin{eqnarray}
-t^{z-y-1} \int_ x^\infty{du \left[ yf\left( u\right) + zuf'\left( u\right)
\right]} =\nonumber\\
 \left[ \dots t^{-zk} x^{-k} - \dots t^{-z\left( 1\right)}\right]
t^{-y} f\left( x\right) .
\label{iiaS1}
\end{eqnarray}
The quasi-static results for small $m$ came 
from achieving a cancellation on the
RHS. For the large $m$ case, the different $x$-dependences preclude
cancellation, but both terms on the RHS have the same dominant order if Eq.
(\ref{z1}) is valid. As shown in Sec. \ref{scalingtheory}, this is consistent
with our scaling theory and with simulation data.

However, the dominant $t$-dependences 
in the LHS and RHS of Eq. (\ref{iiaS1})
give
\begin{equation}
z-y-1 = -zk-y \qquad\Rightarrow\qquad z=\frac{1}{k+1} .
\label{ziia}
\end{equation}
Comparison with $z=\frac{1}{k+2}$, given by Eq. (\ref{Mt}) and confirmed by
simulation, shows that this result is not correct. The origin of the
discrepancy is an important correlation missed by the IIA, as will be discussed
below.

\subsection{Inadequacy of the IIA and a heuristic adjustment}
\label{inadequacy}

The temporal evolution at large $m$ is being misrepresented by the IIA because
it associates a product weight $P_t\left( m\right)P_t\left( 1\right)$ to the
joint occurrence of a free particle and a cluster of mass $m$, not
distinguishing between cases where the particle and the cluster are adjacent or
well separated. These two cases are very different for large $m$ because of the
small probability of detachment of a particle from a large cluster and the high
probability of the subsequent random walk of the particle finishing with
absorption at the originating cluster. In the full original Master equation
[Eq. (\ref{mastergeral}), with cluster/column labels and without factorisation
of probabilities] it is easy to identify the random walk steps (through the  
$i$-labels, and since they occur with rate $\gamma (1)$). For comparison, we can
also see them through $A_t\left( m,l\right)$ in the IIA version retaining
column labels 
[Eqs. (\ref{master1}), (\ref{defA}), (\ref{defA1}) and (\ref{defJ})],
where here the inadequate factorisation has been made (which does not properly
represent the distortion of the walk by the large cluster).

These problems can be adjusted as follows. The absorbing aspect of the random
walk of a single particle near a large cluster reduces the effective rate of
migration to another large cluster. Given that their average separation
increases as their average size, $t^z$, the reduction is by an extra factor
$t^{-z}$, to be introduced into the terms on the RHS of Eq. (\ref{iiaS})
(consequently, the RHS of Eq. \ref{iiaS1} also changes by the extra factor
$t^{-z}$). This
is equivalent to the effect included in the scaling arguments of Sec.
\ref{scalingtheory}. The consequence is that in place of Eq. (\ref{ziia}), the
power counting gives
\begin{equation}
z-y-1 = -zk-y-z \qquad\Rightarrow\qquad z=\frac{1}{k+2} .
\label{ziiacor}
\end{equation}
The extra factors $t^{-z}$ do not modify the
quasi-static form for the distribution function $P_t (m)$ at small $m$, thus
its introduction is still consistent with Eq. (\ref{z1}).

Thus, using these heuristic arguments we are able to predict the correct
coarsening exponent and preserve several exponents relations. However, the
changes are still unable to predict the logarithmic corrections shown in Sec.
\ref{scalingcoars}, which are related to a crossover from symmetric to
asymmetric particle exchange between neighboring clusters.

\subsection{Cluster size distributions}

Using Eqs. (\ref{z1}) and (\ref{Gammat}), the quasistatic result (\ref{ptm}) for
small masses can be rewritten as
\begin{equation}
P_t\left( m\right) \sim {\left( \frac{m}{t^z}e\right)}^{mk} .
\label{ptm1}
\end{equation}
Comparing with $t^{-y} f\left(\frac{m}{t^z}\right)$ (Eq. \ref{scalingP}), it can
be estimated that the crossover between the forms for the region A and the
scaling region S occurs at $m=m_0 (t)$ where
\begin{equation}
m_0 (t) \sim \frac{1}{e} t^z \left[ 1 + {\cal
O}\left( t^{-z}\ln{t}\right)\right] .
\label{m0t}
\end{equation}
The form (\ref{ptm1}) first decreases with $m$ (due to the increasing power of
$t^{-z}$) but then turns over into an increasing function when $m$ exceeds
${\cal O}\left( t^z\right)$. 
The minimum is at $m=\overline{m}\left( t\right)$
such that $0=\frac{d}{dm}\left[ \ln{P\left( m\right)}\right] \sim
\frac{d}{dm}\left[ mk\left( \ln{m} -z\ln{t} -1\right)\right] =
k\left( \ln{m} -x\ln{t}\right)$, so
\begin{equation}
\overline{m}\left( t\right) = t^z .
\label{mmt}
\end{equation}
Thus the minimum is near the crossover region.
Simulations consistently show that the scaling starts just beyond the minimum
and that the quasistatic results (\ref{ptm}) and (\ref{ptm1}) work well up to
just beyond the minimum.

Eqs. (\ref{Mt}) and (\ref{mmt}) imply that the position of the minimum
decreaases with increasing $k$. This is also seen in simulations, and is
consistent with small clusters having largely $m=1$ for $k>1$ and a greater
spread for $k<1$.

The adjusted form of Eq. (\ref{iiaS1}) for the scaling function (Sec.
\ref{inadequacy}) is, using Eqs. (\ref{z1}) and (\ref{yz}), 
\begin{equation}
-\int_x^\infty{du \left[ 2f\left( u\right) + uf'\left( u\right) \right]} =
\left( ax^{-k} - b\right) f\left( x\right)
\label{fadj}
\end{equation}
where the factors of $t$ have consistently cancelled by using the correct
coarsening exponent (Eq. \ref{ziiacor}), 
and $a$ and $b$ are constants
associated with $\gamma (m)$ and $\Gamma$, respectively. Differentiating Eq.
(\ref{fadj}) with respect to $x$ gives $-\frac{d\ln{f\left( x\right)}}{dx} =
\frac{2-akx^{-\left( k+1\right)}}{x+ax^{-k}-b}$, hence
\begin{equation}
f\left( x\right) \propto \exp\left[
-\int{dx \frac{2-akx^{-\left( k+1\right) }}{x+ax^{-k}-b} } \right] .
\label{fx}
\end{equation}

For small $x$, the integrand in the indefinite integral is dominated by
$-kx^{-\left( k+1\right)} /x^{-k}$, which integrates to $\ln{x^{-k}}$, thus
\begin{equation}
f\left( x\right) \propto x^k .
\label{fx1}
\end{equation}
For large $x$, the dominant part of the integrand is $2/x$, giving
\begin{equation}
f\left( x\right) \propto x^{-2} .
\label{fx2}
\end{equation}

\section{Simulation results for cluster size distributions}
\label{distr}

Despite the problems of the IIA to predict the coarsening exponents and the
absence of the logarithmic corrections in the time scaling, even after suitable
adjustment (Sec. \ref{inadequacy}), it progresses beyond the previous scaling
theory (Sec. \ref{scalingtheory}) by providing information on the cluster size
distributions, which can now be compared to simulation data.

The unusual shape of the cluster size distribution in this problem is
illustrated in Fig. 7 for $k=1$ ($t=5\times {10}^5$) and $k=0.5$ ($t={10}^6$). 
There is a rapid (power-law) decrease of $P(m)$ for small $m$, usually until
$m$ of order $10$, and a peak appears at large $m$, i. e. in the range of
typical large clusters. For $k>1$, the statistical weight of the small clusters
decreases with time, i. e. the left side of the curve becomes smaller when
compared to the peaked region. For $k<1$ the opposite occurs: as time
increases, the weight of the small $m$ region increases and the peak becomes
relatively smaller. Indeed, the curve for $k=1/2$ in Fig. 7 shows that the
probability of
isolated particles or dimers is $100$ to $1000$ times larger than the
probability of sizes in the peaked region [for instance, $P(1)\approx 0.77$].

\begin{figure}[!h]
\includegraphics[clip,width=0.4\textwidth,angle=0]{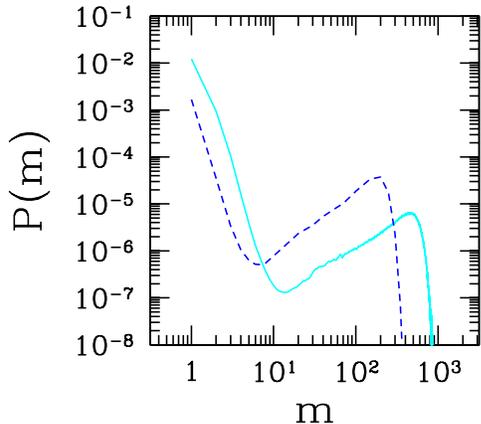}
\caption{(Color online) Cluster size distributions for $k=1$ at $t=5\times
{10}^5$ (dashed curve) and $k=0.5$ at $t={10}^6$ (solid curve). }
\label{fig8}
\end{figure}

The first important result of the IIA is Eq. (\ref{scalingP}) for the scaling
region (the region of the peak in Fig. 7),
with $y$ given by Eq. (\ref{yz}). Simulations show that this result is valid
with $t$ replaced by $t_l=t/\ln{t}$, which is an expected correction. This is
illustrated in Figs. 8a and 8b, where we show
$\log{\left[ t_l^yP_t\left( m\right)\right]}$ as a function of $m/{t_l}^z$ for
$k=0.5$ and $k=2$, respectively, and three different times for each $k$. The
good data collapse (particularly for the largest times) is obtained with
$z=1/\left( k+2\right)$ and $y=2z$, as predicted by Eqs. (\ref{Mt}) and
(\ref{yz}).

A power-law in the left tail of the scaling function $f(x)$ is observed in our
simulations, but the exponents are different from those predicted in Eq.
(\ref{fx1}) for small $k$. For instance, for $k=0.5$, the exponent is $1.07$.
For larger $k$, the agreement is slightly better, e. g. exponent $2.05$ for
$k=2$. Anyway, one interesting feature of the IIA results (\ref{fx1}) and
(\ref{fx2}) is that the left tails of the distributions are heavier than their
right tails for $k<2$. In other words, the distributions have negative
skewness. This is clearly observed in Fig. 8a, for $k=0.5$, while for $k=2$
(Fig. 8b) the skewness is closer to zero (but still negative).

\begin{figure}[!h]
\includegraphics[clip,width=0.4\textwidth,angle=0]{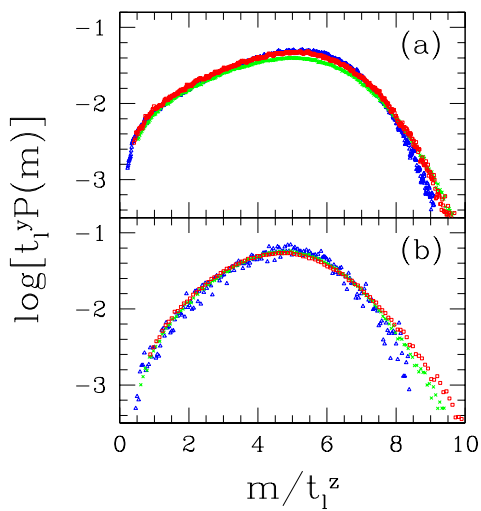}
\caption{(Color online) Scaled cluster size distributions in the scaling region
for: (a) $k=0.5$ at $t={10}^5$ (crosses), $t=2\times {10}^5$ (squares) and
$t={10}^6$ (triangles); (b) $k=2$ at $t=2\times {10}^5$ (squares), $t={10}^6$
(crosses) and $t=5\times {10}^6$ (triangles).}
\label{fig9}
\end{figure}

The negative skewness of cluster size distributions is an uncommon feature in
this type of problem in one dimension; for instance, the distributions in
coarsening with constant detachment rates are positively skewed
\cite{chamereis}, as well as those in the steady states with some rate
functions which increase with cluster size (due e. g. to repulsive
interactions) \cite{gama}. Thus, in a real system, that feature would suggest
the presence of attractive interactions leading to a decrease of the
detachement rate with cluster size. On the other hand, it is important to
notice that it is a common feature in two dimensions, both in point islands
models (which are two-dimensional ZRP) and in extended islands models
\cite{etb}. 

The scaling of small masses (Eq. \ref{ptm1}) is confirmed in Figs. 9a and 9b
for the same values of $k$, again with the logarithmic corrections in the time
$t$. There we plot  $\log{\left[ P_t\left( m\right)\right]}$ versus
$m\log{\left( m/t^z\right)}$, which is proportional to the logarithm of the
RHS of Eq. (\ref{ptm1}). In Figs. 9a and 9b, one important point is the large
range of both variables (horizontal and vertical), which span 2 to 6
orders of magnitude. This explains the discrepancies from a perfect data
collapse when compared to the scaling regime in Figs. 8a and 8b.

\begin{figure}[!h]
\includegraphics[clip,width=0.4\textwidth,angle=0]{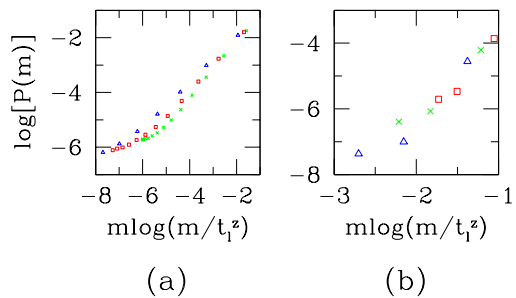}
\caption{(Color online) Scaled cluster size distributions in the small mass
region for(a) $k=0.5$ and (b) $k=2$. Symbols are the same of Figs. 9a and 9b.}
\label{fig10}
\end{figure}

\section{Conclusion}
\label{conclusion}

We studied conserved one-dimensional models of particle diffusion, attachment
and detachment from clusters, where the detachment rates decrease with
increasing cluster size as $\gamma (m)\sim m^{-k}$. Heuristic scaling arguments
based on random walk properties were used to predict the scaling of the typical
cluster size as ${\left( t/\ln{t}\right)}^z$, with $z=1/\left( k+2\right)$. The
coarsening of neighboring clusters is characterized by initial symmetric flux
of particles between them followed by an effectively assymmetric flux due to
the unbalanced detachment 
rates (despite the symmetric model rules). For
$k<1$, the average cluster size does not scale as the size of typically large
clusters due to the high densities of small clusters, which dominate that
average. We also solve the Master equation of the model under an independent
interval approximation, which predicts some exponent relations and the correct
dominant coarsening exponent after suitable changes to incorporate effects of
correlations. These results are confirmed by simulation, which also shows the
negatively skewed cluster size distributions (particularly for small $k$) and
the different scaling relations followed by small clusters (sizes of order $1$)
and by typically large clusters (size of order $t^z$).

The rate functions analyzed here may arise from associating (Arrhenius)
detachment rates with potentials $U(m)$ for particles at the end of a cluster
of size $m$. $U(m)$ is then a sum, from $l=1$ to $m-1$, of pair potentials
$V(l)$ for separation $l$ with Coulomb-like (inverse of distance) attractive
form. In a real system, such interaction is not expected to be valid for all
sizes, but may be a reasonable approximation for some ranges, in a similar way
that long range repulsion between adatoms on a surface represents
substrate-mediated interactions. The particular coarsening features discussed
here will certainly help to identify such application.

It is also interesting to note that our model with $\gamma(m)\sim
\exp{\left( -m^2\right)}$ was already studied in Ref. \protect\cite{granular}
and quantitatively describes experiments with a shaken "gas" of steel beads
distributed among a set of boxes. The average cluster size increases as
${\left( \log{t}\right)}^{1/2}$ and the density of particles in the boxes
without big clusters decrease as $1/t$. These results can be obtained by a
direct extension of the scaling arguments of Sec. \ref{scalingtheory} (the
second one may be viewed as the $k\to\infty$ limit of Eqs. \ref{rhofree} and
\ref{z1}).

From the theoretical point of view, this work contains some
important advances. First, we show how random walk properties and simple model
rules are able to predict the coarsening law including a logarithmic
correction, which is a non-trivial task at the level of a scaling theory.
Moreover, this correction is shown to be a consequence of a continuous
competition between symmetric particle flux between neighboring clusters and a
dominant assymmetric flux, despite the absence of a
spatial bias in the model rules, in contrast with other ZRP where 
asymmetric flux appeared only
as a consequence of such bias. Finally, the different scaling relations obeyed
by small clusters and by typically large clusters, which enable the former to
be statistically dominant when $k<1$, contrasts with other models with similar
physical mechanisms (even those involving deposition and/or fragmentation),
where a single scaling relation is sufficient to represent all relevant cluster
sizes.

\begin{acknowledgments}

FDAA Reis thanks the Rudolf Peierls Centre for Theoretical Physics of Oxford
University, where this work was done, for hospitality, and acknowledges
support by the Royal Society of London (UK) and Academia Brasileira de
Ci\^encias (Brazil) for his visit.

RB Stinchcombe acknowledges support
from the EPSRC under the Oxford Condensed Matter Theory Grants,
numbers GR/R83712/01, GR/M04426 and EP/D050952/1.

\end{acknowledgments}

\end{document}